\documentclass[aps,prl,twocolumn]{revtex4}
\usepackage{graphicx}% Include figure files
\usepackage{bm}% bold math

\begin{document}
\def\be{\begin{equation}} 
\def\ee{\end{equation}}
\def\bearr{\begin{eqnarray}}
\def\eearr{\end{eqnarray}}
\def\tc{$T_c~$}
\def\tcl{$T_c^{1*}~$}
\def\c2{ CuO$_2~$}
\def\ruo{ RuO$_2~$}
\def\lsco{LSCO~}
\def\bi{bI-2201~}
\def\tl{Tl-2201~}
\def\hg{Hg-1201~} 
\def\sro{$\rm{Sr_2 Ru O_4}$~}
\def\rc{$RuSr_2Gd Cu_2 O_8$~}
\def\mgb{$MgB_2$~}
\def\pz{$p_z$~}
\def\ppi{$p\pi$~}
\def\sqo{$S(q,\omega)$~}
\def\tperp{$t_{\perp}$~}
\def\he4{${\rm {}^4He}$~}
\def\ags{${\rm Ag_5 Pb_2O_6}$~}
\def\nxcob{$\rm{Na_x CoO_2.yH_2O}$~}
\def\lsco{$\rm{La_{2-x}Sr_xCuO_4}$~}
\def\lco{$\rm{La_2CuO_4}$~}
\def\lbco{$\rm{La_{2-x}Ba_x CuO_4}$~}
\def\half{$\frac{1}{2}$~}
\def\thalf{$\frac{3}{2}$~}
\def\tst{${\rm T^*$~}}
\def\tch{${\rm T_{ch}$~}}
\def\jeff{${\rm J_{eff}$~}}
\def\nbc{${\rm LuNi_2B_2C}$~}
\def\cabc{${\rm CaB_2C_2}$~}
\def\nboo{${\rm NbO_2}$~}
\def\voo{${\rm VO_2}$~}
\def\nip{$\rm LaONiP$~}
\def\fep{$\rm LaOFeP$~}
\def\cop{$\rm LaOCoP$~}
\def\mnp{$\rm LaOMnP$~}
\def\efeas{$\rm LaO_{1-x}F_xFeAs$~}
\def\hfeas{$\rm La_{1-x}Sr_xOFeAs$~}
\def\hSfeas{$\rm Sm_{1-x}Sr_xOFeAs$~}
\def\hCefeas{$\rm Ce_{1-x}Sr_xOFeAs$~}
\def\feas{$\rm LaOFeAs$~}
\def\Ndfeas{$\rm NdOFeAs$~}
\def\Prfeas{$\rm PrOFeAs$~}
\def\ttog{$\rm t_{2g}$~}
\def\eg{$\rm e_{g}$~}
\def\dxy{$\rm d_{xy}$~}
\def\dzx{$\rm d_{zx}$~}
\def\dzy{$\rm d_{zy}$~}
\def\dxsq{$\rm d_{x^{2}-y^{2}}$~}
\def\dzsq{$\rm d_{z^{2}}$~}

\title{Quantum String Liquid State in \feas and Superconductivity}

\author{ G. Baskaran \\
Institute of Mathematical Sciences\\
C.I.T. Campus,
Madras 600 113, India }

\begin{abstract}
Superconducting \feas family of even electron metallic systems have striking resemblance to odd electron cuprates. We suggest that this resemblance is caused by presence of two coupled 2D resonating valence bond systems in \feas. Bond charge repulsion and Hund coupling fuse the 2 species of valence bonds into closed strings with a Haldane gap, resulting in a quantum string liquid. A pair of doped holes (electrons) creates an open string and remain at the ends as holon (doublon). Charge $\pm$ 2e singlet strings condense to produce high Tc superconductivity. Higher Tc's are likely in our \textit{string route to High Tc superconductivity}, when competing orders are taken care of.
\end{abstract}

\maketitle

Progress in material science and technology continues to generate novel many body quantum states. In the past decades we have witnessed the birth of quantum Hall states, heavy fermions, RVB superconductors, quantum spin liquids, orbital liquids etc. In the wake of these progress, a wealth of novel ideas and theoretical techniques, with far reaching implication to basic science and technology have emerged. Recently superconductivity in \fep with a \tc $\approx$ 5 K was discovered\cite{fep}, followed by an exciting discovery\cite{feas} of superconductivity with a \tc $\approx$ 28 K in electron doped \efeas. Following heels, hole doped \hfeas has been synthesised\cite{hfeas} with a \tc $\approx$ 25 K, making electron and hole doping nearly symmetric, reminding us of particle-hole symmetric one band Hubbard model physics. More recently\cite{Ndfeas,Prfeas}, replacement of La by other rare earths Sm, Ce, Nd and Pr have resulted in a substantial increase in Tc, 52 K being the highest so far.

The even electron character of Fe$^{2+}$ in \feas, orbital degeneracy and observed large superconducting \tc makes it an unusual high Tc superconductor. Is nature unfolding another novel organisation of strongly correlated electrons in this 3d$^{6}$ system ? In this letter we suggest formation of a novel \textit{quantum string liquid}(QSL). The beads of the strings are two types of valence bonds of two spin-\half resonating valence bond (RVB) systems available at the fermi level in \feas. Fusion of valence bonds into strings arise from mutual bond charge repulsion of two different types of overlapping valence bonds and Hund coupling. The closed singlet strings have a Haldane type correlation and a spin gap\cite{haldane}

After a brief introduction we write down a model and present an RVB mean field theory. Going beyond meanfield theory involving 3 different projections brings out the string content of the new superconductors. Our estimate of superconducting Tc gives a large value in the range 100 to 200 K. We then briefly indicate how the underlying string structure can bring about real space orders that will in general compete with superconductivity.

The layered transition metal monopnictides , $\rm{LaOTX}$ (T = Mn, Fe, Co Ni; and  X = P, As) exhibit a variety of low temperature phases: \fep and \nip are both superconductors\cite{nip} with a low \tc $\approx$ 4 and 6 K. \cop is a ferromagnetic metal with a \tc $\sim$ 50 K and \mnp is an insulator\cite{feas}. Further, iron monopnictides offer some surprise: stoichiometric \fep is a superconductor, whereas \feas becomes a superconductor only after doping. The spin susceptibility of \feas is Pauli like but considerably enhanced compared to cuprates. There is a resistivity anomaly in \feas at about 150 K. Transport, specific heat, $\rm {H_{c2}}$ and tunnelling measurements have been reported. The hall coefficients of electron doped \efeas is negative\cite{hall}. In an exciting work\cite{hfeas}, Sr doped \hfeas has been synthesised, resulting in a hole doped superconductor with a \tc $\approx$ 25 K. The hall constant of \hfeas is positive. The superconducting Tc has steadily increased from 26 K all the way up to $\approx$ 52 K in doped \Prfeas and \Ndfeas. Recent experiments have also confirmed a suggested spin density wave order\cite{OakRidge1}.

From theory point of view, the first electronic structure calculation was performed by Lebegue\cite{lebegue} for superconducting \fep. Following the discovery of high \tc superconductivity in \efeas, several groups have performed LDA and more sophisticated calculations\cite{LDAtheories} for undoped and doped \feas and also have discussed the importance of strong electron interactions in this d-band system. All these calculations suggest a dominantly non-phononic mechanism of superconductivity.  Nature of Spin density wave order in \feas has also been discussed. Different mechanisms and order parameters have been also proposed for the superconducting state\cite{mechanisms}.

\feas is a layered material with alternate stacking of LaO and FeAs layers. For modelling purposes it is important to remember that Fe atoms form a square lattice, a quasi 2 dimensional metal. Each Fe atoms is tetrahedrally coordinated by As atoms. 

Fe$^{2+}$ in \feas is in 3d$^{6}$ configuration. LDA calculations show 5 overlapping 3d bands spread in the energy range -2 to +2 eV. One sees three filled bands and 2 empty bands, that have a small overlap. The overlap leads to small fermi surface pockets with fermi energy $\sim$ 0.2 eV. We suggest that this is not a good starting point to build low energy physics; we have to take care of strong correlation effect. We find that strong correlations splits the 6  electron \feas bands into a pair of spin-\half resonating valence bond systems close to the fermi level, as shown schematically in figure 1. We call it as 2 RVB system. We will present this analysis in a separate communication. 

\begin{figure}[h]
\includegraphics[height=1.8in]{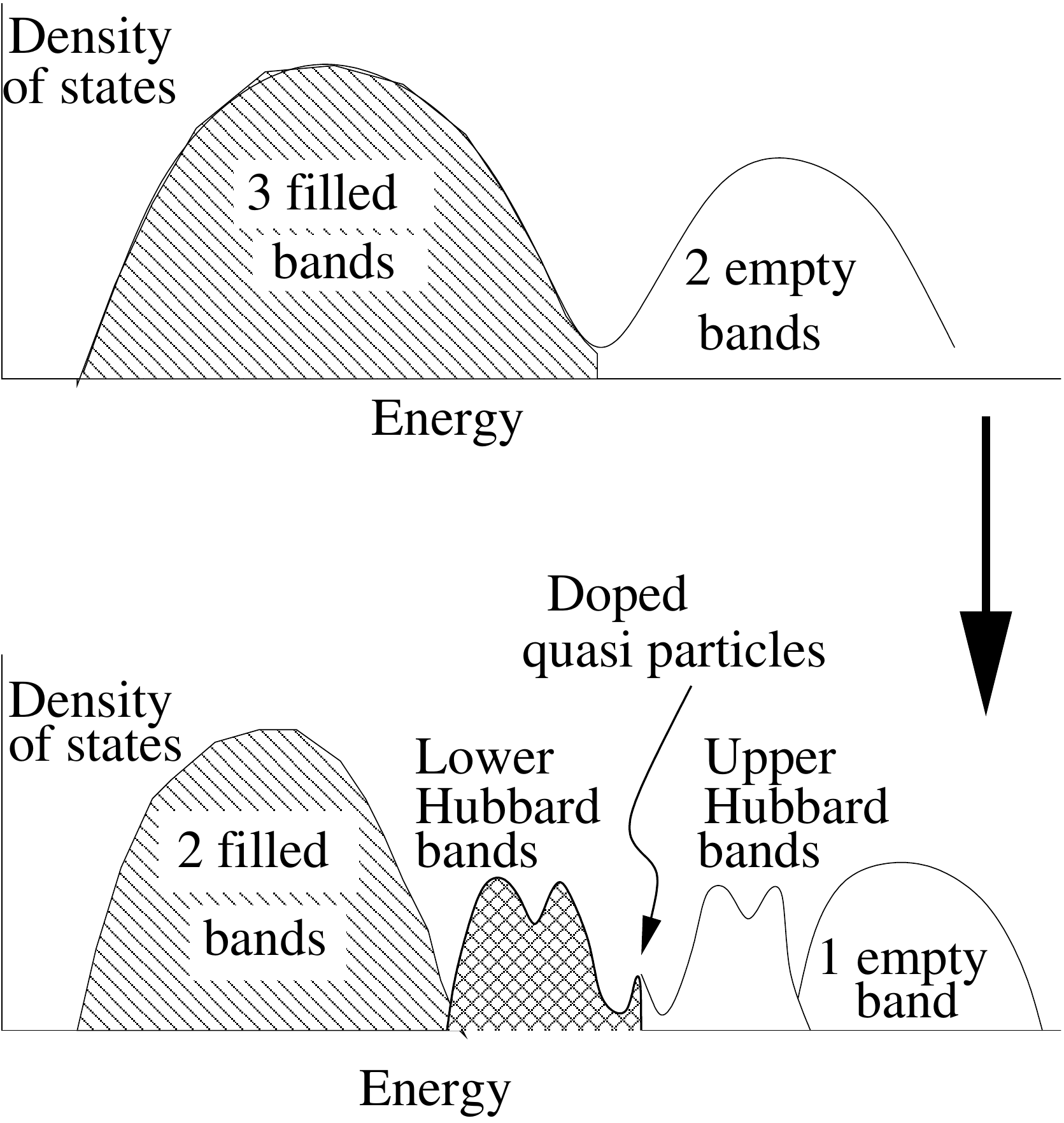}
\caption{Schematic picture of splitting off of two half filled bands at the fermi energy, from filled bands as coulomb repulsion is increased.}
\end{figure}

Our starting point of this paper in discussing mechanism and theory of superconductivity in doped \feas will be a 2 RVB system. A major prediction of our model building is that ARPES should see two large fermi surfaces close to half filling, rather than small fermi surface pockets predicted by LDA calculations.

LDA calculations finds that the tetrahedral crystal field splitting and a large direct 3d orbital overlap spread the spectral weight of every 3d orbital over the entire band width of 4 eV. As for the symmetry of the Wannier orbitals of these two bands are concerned, band order suggested by tetrahedral crystal field splitting need not be a good guidance, because direct Fe 3d orbital overlap produce orbital mixing and compete with crystal field splitting. Based on symmetry grounds Wannier orbitals will have the same symmetry property of the 3d orbitals; however, they will have strong admixture with various d orbitals of neighboring sites. 

Assuming that the metallic state maintains the square lattice symmetry, there are different possibilities for choosing symmetry of the two Wannier orbitals: i) a pair from 3\dxsq, 3\dxy, 3\dzsq, whose $|\psi|^{2}$ have square planar symmetry` or ii) 3\dzx and 3\dzy. The second choice will give 2 spatially anisotropic bands, but together will maintain the square lattice symmetry. 

So we choose a fairly general two orbital Hamiltonian. Key parameters of our model are inter and intra orbital Hubbard U, $\sim $ 3 to 4 eV, bond charge repulsion V$_{12}\sim$ 1 eV and Hund coupling $\sim$ 0.7 - 1.0  eV. Width of an individual 3d band is about 2 eV, giving a hopping parameter between nearest neighbour Wannier orbitals, t $\sim$ 0.5 eV.

The Hamiltonian of our 2 RVB systems is a two orbital Hubbard model:
\bearr
&&H =  -\sum_{ij\mu} t^{}_{ij\mu}c^{\dagger}_{i\mu \sigma}c^{}_{j\mu \sigma} + h.c. + 
\sum_{i \mu} U_{\mu}n_{i\mu \uparrow}n_{i\mu \downarrow} + \nonumber \\
&& + V_{\rm 12}\sum_{\langle ij \rangle} n_{ij1} n_{ij2} 
- J_{H} \sum_{i} (\sum_{\mu}c^{\dagger}_{i\mu \alpha} {{\vec \sigma}}_{\alpha \beta}c^{}_{i\mu \beta})^2 
\eearr
Here $\mu, \nu = 1,2$ represents Wannier orbitals. Hopping is assumed to exist only among same type of nearest neighbour orbitals. Second line of the above equation couples the two systems, through bond charge repulsion and Hund coupling.
The operator n$_{ij\mu} \equiv \frac{1}{2}\sum_{\sigma} (c^{\dagger}_{i\mu \sigma} + c^{\dagger}_{j\mu \sigma})
(c^{}_{i\mu \sigma} + c^{}_{j\mu \sigma}) $ counts number of electrons in bonding state of $\mu$-th orbital connecting neighboring sites i and j. The bond charge repulsion is a coulomb interaction term,V$_{12} = $$\int |\psi_{i1}({\bf r})+\psi_{j1}({\bf r})|^2\frac{e^{2}}{|{\bf r - r'}|}
|\psi_{i2}({\bf r'})+\psi_{j2}({\bf r'})|^2 d{\bf r} d{\bf r'}$, where $\psi_{i\mu}$ are the two Wannier orbitals at site i.

Starting from the above Hamiltonian, using a superexchange perturbation theory, we derive the following effective Hamiltonian for the optimally doped case. It is a sum of two t-J models:
\bearr
&& H_{\rm{eff}} \equiv H_{\rm tJ1} + H_{\rm tJ2} = -\sum_{ij\mu} t^{}_{ij\mu}c^{\dagger}_{i\mu \sigma}c^{}_{j\mu \sigma} + h.c.+ \nonumber \\
&&- \sum_{ij}J_{ij\mu} ({\bf S_{i\mu}} \cdot {\bf S_{j\mu}} - \frac{1}{4}n_{i\mu}n_{j\mu})
\eearr
with three local constraints: i) $n_{i\mu \uparrow}+ n_{i\mu \downarrow} \neq 0~{\rm or}~2$, for electron or hole doped cases, ii) $\sum_{\mu} b^{\dagger}_{ij\mu}b^{}_{ij\mu} \neq 2$
and iii) $(\sum_{\mu}c^{\dagger}_{i\mu \alpha} {{\vec \sigma}}_{\alpha \beta}c^{}_{i\mu \beta})^2  \neq 0 $. The superexchange J $\approx \frac{4t^{2}}{U}$. At the Hamiltonian level the 2 RVB systems are decoupled; however the three constraints couple two RVB systems in a non trivial fashion. The first, double/zero occupancy constraint is well known in t-J model. The second and third are new in the context of superconductivity theory. The second one tells us that two neighboring sites containing 2 electrons each cannot form two covalent bonds, because of the bond charge repulsion. The third constraint is dictated by Hund coupling, which favours maximal spin at a given site. 

We will discuss the above model Hamiltonian, using RVB theory approach\cite{pwa,bza}: i) solve the unconstrained Hamiltonian in a mean field theory and ii) perform all projections in the resulting wave function, for further analysis. The unconstrained Hamiltonian is the same as the one solved in the first RVB mean field theory\cite{bza} of cuprates. Later several important improvements\cite{vanilla} have been made on that approach. We can use all those results. Formally the wave function we wish to analyse for superconductivity in our 2 RVB system (for hole density x) is the following:
\bearr
|SC\rangle \approx P_{G} P_{B} P_{S}&& \left( \sum_{ij} \phi_{ij} b^{\dagger}_{ij1}\right)^{\frac{N}{2}(1-\frac{x}{2})} \times \nonumber \\
\times &&\left( \sum_{ij} \phi_{ij} b^{\dagger}_{ij2}\right)  ^{\frac{N}{2}(1-\frac{x}{2})} |0\rangle \nonumber \\
 \equiv P_{G} P_{B} P_{S}&& \left( \sum_{ij\mu} \phi_{ij} b^{\dagger}_{ij1\mu}\right) ^{\frac{N}{2}(1-x)}|0\rangle
\eearr
where $b^{\dagger}_{ij\mu} = \frac{1}{\sqrt{2}}(c^{\dagger}_{i\mu \uparrow}c^{\dagger}_{i\mu \downarrow}
- c^{\dagger}_{i\mu \downarrow} c^{\dagger}_{i\mu \uparrow})$ is the singlet operator. Further, P$_{G}$ is the usual Gutzwiller projection which prevents double occupancy in any of the orbitals. P$_{B}$ avoids avoids two singlet bonds connecting two neighboring sites. P$_{S}$ projects out the singlet spin component of two electrons at any site. The pair function $\phi_{ij}$ is the variational wave function. For a single RVB system, in a square lattice, the above function has \dxsq symmetry and leads to a superconducting state with nodal quasi particles. 
\begin{figure}[h]
\includegraphics[height=1.6in]{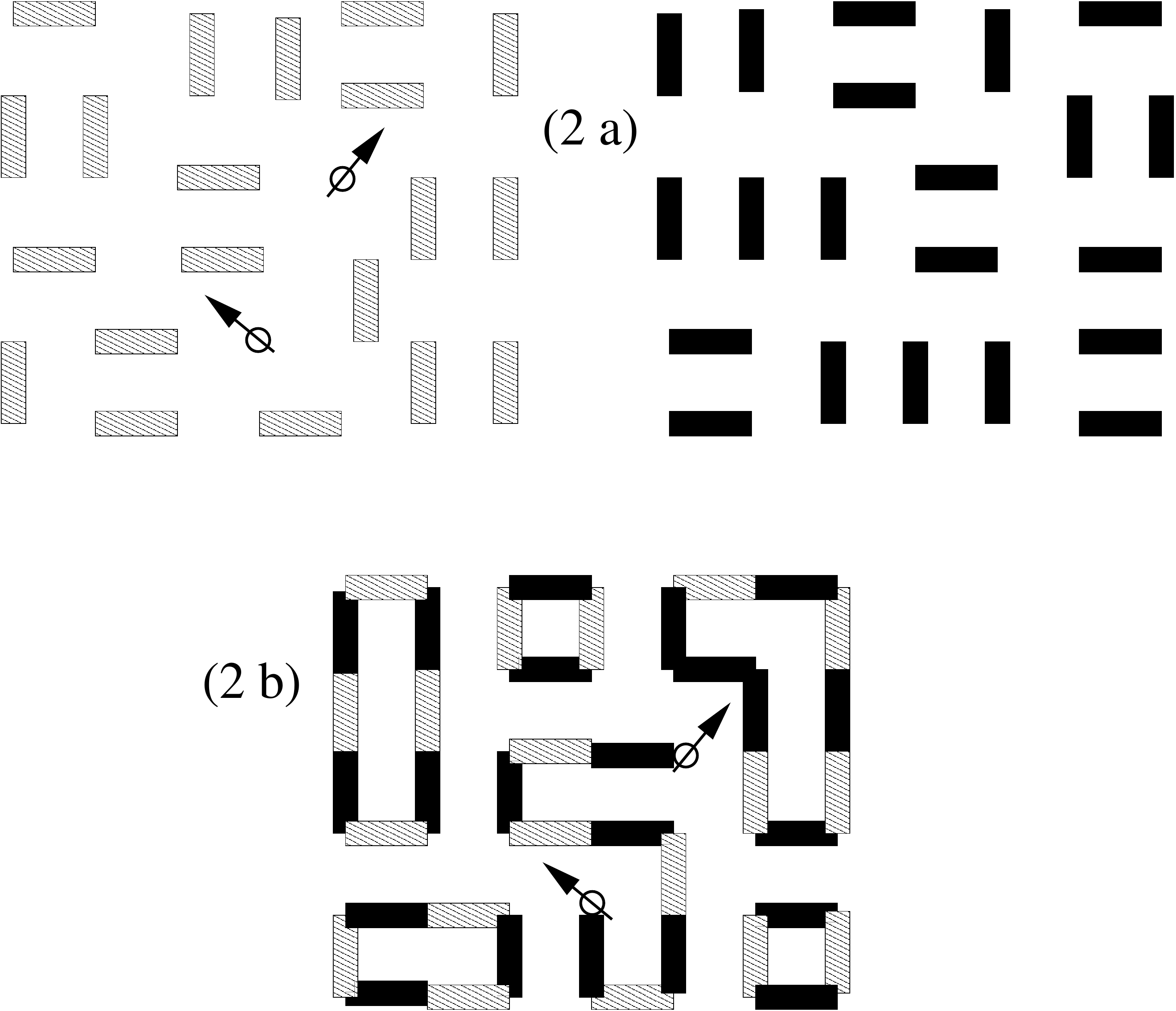}
\caption{Fusion of Valence Bond States. 2a) VB states of the two systems. 2 b) String like organization after fusion.}
\end{figure}
In the presence of coupling between two RVB systems key questions are: i) do the three projections preserve d-wave superconductivity of a single RVB system? and ii) are there going to be some new physics? These are difficult questions to answer. However, the available phenomenology suggests superconductivity, similar to that suggested by 2D repulsive Hubbard model close to half filling survives. In what follows, we will address these question from the point of view of short range RVB states and find that projections bring an entirely new quantum liquid namely quantum string liquid and associated rich possibilities.

We also develop a quantum mechanical basis of singlet string states, by an appropriate fusion of two short range RVB states. We will start with the case of no doping. Consider two nearest neighbour valence bond (VB) states shown in figure 2a. The VB's of two systems are denoted by dark and shaded bonds. If we choose two arbitrary VB states and fuse them, we will get overlapping bonds, in general. Such states are energetically not favourable because of bond charge repulsion energy V$_{12}$. The two valence bond states in figure 2a do not have bond overlap. When we fuse them we get a state shown in 2b. From simple topological considerations it follows that the resulting state is a set of closed strings. Open strings, if they appear, carry spinons at their ends. Also every configuration of closed and open strings (made of nearest neighbour bonds) that fill the lattice can be obtained by a fusion of two unique VB states.

Short range VB states form a overcomplete set of states to describe the physics of a spin-\half Heisenberg model. The overcompleteness make different VB states linearly independent and not orthogonal. As our fused states are direct product of two RVB states, we can use known results of single VB overlap properties to study our combined system.  It should be remembered that the total number of string states is simply not square of the number of possible valence bond states, because of the constraints. 

Let us calculate energy expectation value of the bond repulsion and Hund coupling term in a string state. By construction bond repulsion energy is zero. Since we have two valence bonds meeting at every site, 
the total spin value at a given site is fluctuating: with probability $\frac{1}{4}$ it has value zero and with probability $\frac{3}{4}$ it has value 1. Thus the average value of the Hund coupling energy is $\frac{3}{4}$th of the maximum possible value 2J$_{H}$. It is possible to gain the extra Hund coupling energy $\frac{1}{4}$2J$_{H}$ 
projecting out the singlet component at every site. Or one can replace every string state formally by the exact ground state of the Haldane gapped nearest neighbor Heisenberg antiferromagnetic spin-1 chains. Our strings already have an orbital order as the beads alternate along the chain. Converting them into Haldane chains brings an additional topological order\cite{denijs}.
\begin{figure}[h]
\includegraphics[height=1in]{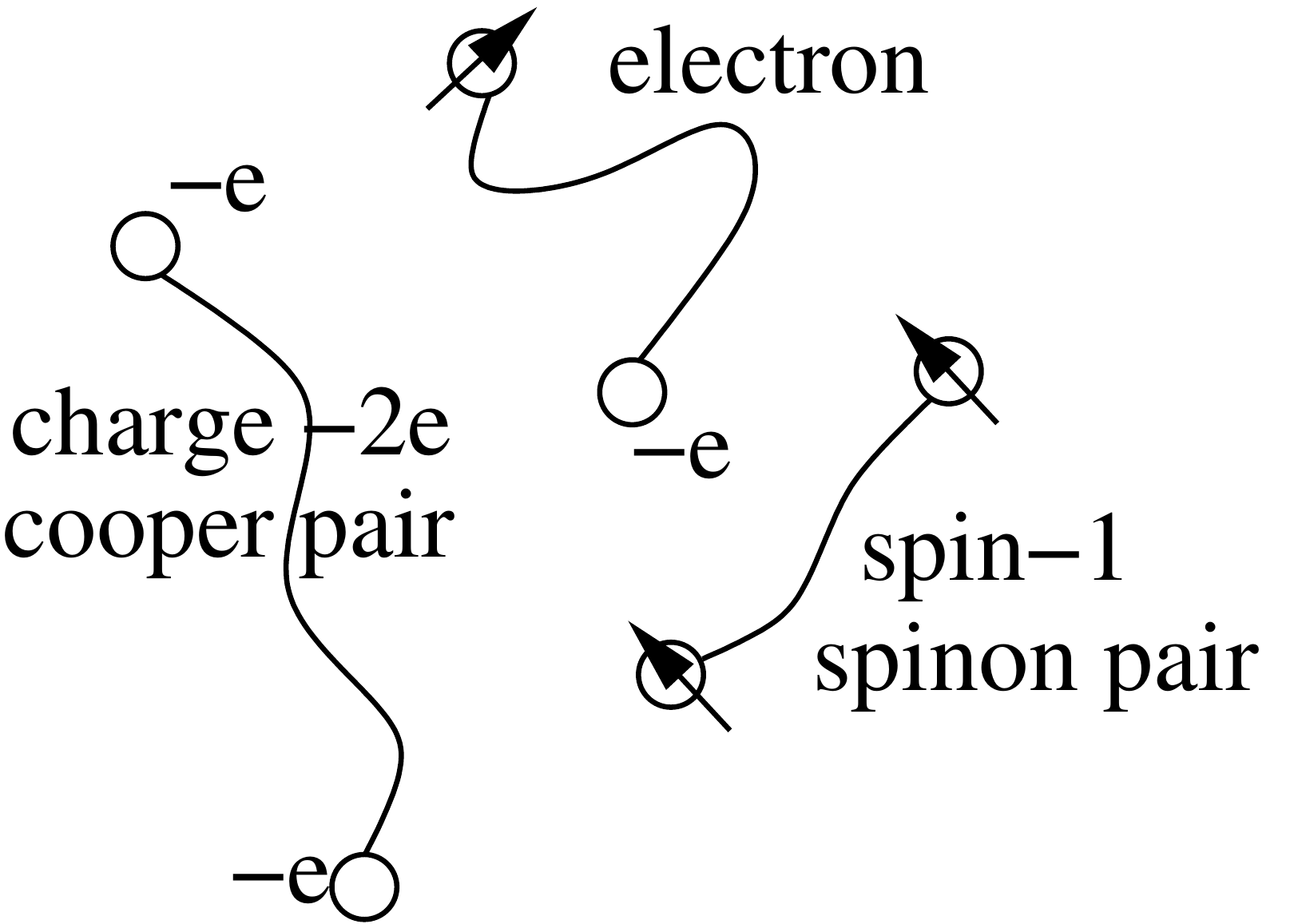}
\caption{charge -2e Cooper pair, a spin-1 spinon pair and an electron, as different string states.}
\end{figure}

Now we briefly discuss topological excitations in our 2 RVB system. Picture 2b already shows how a two spinon (spin-1) state appears as an open string. As every unpaired spin or doublon or holon can occur in one of the two orbitals of a given site, they carry an orbital quantm number as well. A bound spinon pair (figure 3) is likely to be a low energy excitation in our metallic system. When we add an electron to the insulating valence bond state, we get a spinon-doublon composite as shown in figure 3. When two electrons are added, it is energetically advantageous to get rid of two unpaired spins. That is, the unpaired spin will disappear as singlets and we will be left with one open string with two doublons (charge -e) at the ends (figure 3). 

Using our fused short range VB states or equivalently the string state basis, we can describe superconducting ground states and excited states. What is interesting is that the fusion with the constraints has resulted in an incompressible liquid of strings. There are interesting questions about how strings quantum fluctuate in space and manage to create a string liquid. This is a non trivial question, as the string states are also non-orthogonal. Even the simple question of evaluation of overlap of 2 string states becomes complicated, if we keep the strings in Haldane gapped state. More remains to be done to take advantage of the string structure of our 2 coupled RVB states. The nature of holon as well as spinons at the end of the open strings are very interesting. It is known that open Haldane chains support spin-\half
excitations at the boundary\cite{white}. However, the gap moment is not strictly localised on the chain end.
The excess spin density oscillates and decays exponentially, giving a characteristic size for the spinon. We expect the same for the holon as well. 

An open string with charges holon/doublon at both the ends is our cooper pair. It is not obvious how it will modify the nature of superconducting state. While the overall ODLRO and phenomenology may resemble the standard BCS superconductor, there may be subtle topological orders and non trivial excitations in our QSL superconductor. It needs to be explored.

String structure suggests, depending on the size of the open string (dictated by factors such as resonance energy, coulomb repulsion etc.) a binding mechanism for our holon pairs. Is this an additional pairing energy ? The final superconducting state will be a coherent superposition of the resonating charge string configurations. A key parameter that determines the superconducting Tc will be the effective mass of the 2e open string. That will give us a Kosterlitz-Thouless type of scale k$_{B}T_{c} \approx \frac{\hbar^{2} n}{2m_{c}}$, where n is the carrier density per unit area.  This expression is very similar to the expression or Tc in RVB theory, suggested by the condenseation of charge valence bonds or holons. Thus we expect a maximum Tc in the range of 160 to 200 K, perhaps exceeding cuprates.

Since we have string like entities, there will be a tendency for them to have liquid crystalline type order, spin order and charge order, encouraged by unscreened long range interaction at low doping and electron lattice coupling. Such real space organization will in general reduce superconducting Tc. These are competing phases, very much like in cuprates. If one can engineer materials, such as Tl or Hg multilayer cuprates, where charge and spin order tendencies are supressed, superconducting Tc's can go higher than 52 K that has been observed so far. 

Experimentally one sees a metallic state in undoped \feas. This is likely to arise from two possibilities: i) the two reference spin-\half Mott insulators having an occupancy of 1 + x$_{0}$ and 1- x$_{0}$ per site; i.e., they are self doped (x$_{0} < < 1$) for energetic reasons and ii) one reference state remains insulating whereas the other remains strongly correlated but metallic. In the second case a commensurate long range order at $(\pi,\pi)$ can easily occur in the metallic band. External doping may selective enter the Mott insulator and we get superconductivity, somewhat contaminated by the first metallic band. 

The spin susceptibility of the undoped metal \feas and \efeas are both high compared to the metallic cuprates. This is likely to arise from the presence of low energy spin-1 excitations in the quantum fluctuating closed strings. 

It is a pleasure to acknowledge Venky Venkatesan, who brought to my attention reference 2, and knowing my fascination for new superconductors, called it a gift.

\textit{Postscript}: As we were completing this manuscript a paper by Tao Li (cond-mat/0804.0536) appeared. It introduces a model of two strongly correlated 3\dzx and 3\dzy bands and presents a mechanism of spin density wave and superconductivity.


\begin{references}
\bibitem{fep}Y. Kamihara, J. Am. Chem. Soc., {\bf 128}, 10012 (2006)
\bibitem{feas}Y. Kamihara et al., J. Am. Chem. Soc. {\bf 130} 3296 (2008)
\bibitem{hfeas}Hai-Hu Wen et al., cond-mat/0803.3021
\bibitem{Ndfeas}Z.A. Ren et al., cond-mat/0803.4234; G. F. Chen et al., cond-mat/0803.4384
\bibitem{Prfeas}Z.A. Ren et al., cond-mat/0803.4283
\bibitem{haldane} F.D.M. Haldane, Phys. Rev. Lett., {\bf 50} 1153 (1983)
\bibitem{nip}T. Watanabe et al., Inorg. Chem., {\bf 46} 7719?7721 (2007)
\bibitem{hall}H. Yang et al., cond-mat/0803.0623
\bibitem{OakRidge1}C. de la Cruz et al., cond-mat/0804.0795; M.A. McGuire et al., cond-mat/0804.0796
\bibitem{lebegue}S. Lebegue, Phys. Rev., {\bf B 75} 035110 (2007)
\bibitem{LDAtheories}D.J. Singh and M.-H. Du, cond-mat/0803.0429; K. Haule et al., cond-mat/0803.1279; G. Xu, et al., cond-mat/0803.1282;Chao Cao et al., cond-mat.0803.3236; Hai-Jun Zhang et al., cond-mat/0803.4487; L. Boeri et al.,
cond-mat/0803.2703; Gang Xu et al., cond-mat/0803.1282; 
\bibitem{mechanisms}K. Kuroki, cond-mat/0803.3325; Xi Dai et al., cond-mat/0803.3982; I.I. Mazin et al., cond-mat/0803.2740;F. Marsiglio and J.E. Hirsch, cond-mat/0804.0002; Tao Li, cond-mat/0804.0536;G. Giovannetti et al., cond-mat/0804.0866; S. Raghu et al., cond-mat/0804.1113
\bibitem{pwa}P.W. Anderson, Science, {\bf 235} 1196 (1987) 
\bibitem{bza}G. Baskaran et al., Sol. St. Commn., {\bf 63} 973 (1987)
\bibitem{vanilla}P.W. Anderson et al., J. Phys. {\bf C 24} R355 (2004); G. Baskaran, Iranian J. Phys. Res.,
{\bf 6} 163 (2006)
\bibitem{denijs} I. Affleck et al., Phys. Rev. Lett., {\bf 59} 799 (1987); M. den Nijs and K. Rommelse, Phys. Rev., {\bf 40} 4709 (1989); S.M. Girvin and D.P. Arovas, Physica Scripta, {\bf T27} 156 (1989)
\bibitem{white} S.R. White, Phys. Rev. Lett., {\bf 69} 2863 (1993) 
\end{references}
\end{document}